\documentclass{article}

\usepackage{times} 
\usepackage{epsf}

\begin{document}
\title{\bf Encapsulation for Practical Simplification
  Procedures\thanks{This work was supported by the 
  Mathematical, Information, and Computational Sciences Division
  subprogram of the Office of  Advanced Scientific Computing Research,
  Office of Science, U.S. Department of Energy, under Contract
  W-31-109-ENG-38.}}
\author{\emph{Olga Shumsky Matlin and
 William McCune}\\[0.2in]
         Mathematics and Computer Science Division\\
         Argonne National Laboratory\\
         Argonne, IL 60439}
\maketitle
\thispagestyle{empty}

\abstract{ACL2 was used to prove properties of two simplification
  procedures.  The procedures differ in complexity but solve the same
  programming problem that arises in the context of a
  resolution/paramodulation theorem proving system.  Term rewriting is
  at the core of the two procedures, but details of the rewriting
  procedure itself are irrelevant.  The ACL2 {\em encapsulate}
  construct was used to assert the existence of the rewriting function
  and to state some of its properties.  Termination, irreducibility,
  and soundness properties were established for each procedure.  The
  availability of the encapsulation mechanism in ACL2 is considered
  essential to rapid and efficient verification of this kind of
  algorithm.}   

\section{Introduction and Problem Description}
\label{sec:problem}

We examine simplification procedures that arise in resolution,
paramodulation, and rewriting systems.  We have a programming problem,
and at an abstract level we have a straightforward procedure to
solve it.  However, our theorem provers (e.g., Otter \cite{McCune94a}) 
are written in C, with lots of hacks and optimizations that impose
constraints that do not fit with our abstract solution.  We have
devised a two-stage procedure intended to have properties similar to
those of the straightforward procedure.  The two-stage procedure obeys
the constraints, but its correctness is not obvious, so we have called
on ACL2 \cite{acl2:book} for assistance.

The following simplification problem is faced by many
resolution/paramodulation style theorem-proving programs.  Suppose we
have a set {\em S} of clauses with the irreducibility property that no
clause in {\em S} simplifies any other clause in {\em S}.  We wish to
add a new set {\em I} of clauses to {\em S} and have the resulting set
be equivalent to {\em S $\cup$ I} and also satisfy the irreducibility
property.  The problem is interesting because, in addition to members
of {\em S} simplifying members of {\em I}, members of {\em I} can also
simplify members of {\em S}, and those simplified members can simplify
other members of {\em S}, and so on. Consider the following procedure,
which we call {\em direct incorporation}.

\begin{verbatim}
    Q = I;
    While (Q) do
        C = dequeue(Q);
        C = simplify(C, S);
        if (C != TRUE)
            for each D in S simplifiable by C
                move D from S to Q;
            append C to S;
\end{verbatim}

In the terminology of our theorem prover Otter, the statement ``C
= simplify(C, S)" corresponds to both forward rewriting and forward
subsumption, and the loop ``for each D ..." corresponds to back
subsumption and back rewriting.  The list {\em I} represents a set of 
clauses derived by some inference rule.

The direct incorporation procedure does not suit our purposes,
however.  The set {\em I} 
can be too large to generate in full before incorporating it into 
{\em S}.  Members of {\em I} will typically simplify many other
members of {\em I}, so we wish to incorporate {\em I} into {\em S} as
{\em I} is being generated.  Furthermore, the set {\em I} is generated
by making inferences from members of {\em S}, and our algorithms and
data structures do not allow us to remove clauses from {\em S} while
it is being used to make inferences. 

Therefore, we use a two-stage procedure, which we call 
{\em limbo incorporation}.  The first stage simplifies members of 
{\em I} and, if they are not simplified to {\em TRUE}, puts them into
a queue {\em L} (called the {\em limbo list}).  The set {\em S} is not
modified by the first stage.  The second stage processes {\em L} until
it is empty.  For each member {\em B} of {\em L}, all clauses in 
{\em S} that can be simplified by {\em B} are removed from S,
simplified by {\em S $\cup$ L}, then appended to {\em L}.  The second  
stage is similar to the direct incorporation procedure except that in 
the second stage, members of the queue being processed have already 
been simplified with respect to {\em S}.  In Otter terminology, the
first stage does forward simplification, and the second stage does
back simplification. 

The direct incorporation procedure and the limbo incorporation
procedure do not necessarily produce the same results because the
simplification operations can happen in different orders and the
simplifiers we use do not necessarily produce unique canonical forms.  

Our goals are to show, for each incorporation procedure, that (1) it
terminates, (2) it produces a set in which no member can simplify
any other member, and (3) the final set {\em S} is equivalent to the
conjunction of {\em I} and the initial set {\em S}.

\section{ACL2 Solution}

The reasoning we need to do is primarily about the order
in which simplification operations occur and the sets of
simplifiers that are applied.  The details of the basic
simplification procedure and of the evaluation procedure
for proving equivalence properties are irrelevant.  Therefore
we have used an ACL2 encapsulation mechanism to assert the existence 
and relevant properties of the simplification and evaluation
functions.

An alternative to using the encapsulation mechanism is to 
fully define the simplification and evaluation functions and then prove
the required properties based on these formalizations.  Term
rewriting, which is at the core of the simplification procedure, is
not a simple algorithm \cite{term-rewriting}, however, and
considerable effort would have been required to establish its
termination and necessary properties.  Formalizing an evaluation
function would have necessitated formalization of first-order logic in
ACL2, as was done in the {\sc Ivy} \cite{ivy} project.  Our
experiences in that project highlighted the difficulties in
implementing a general first-order evaluation function in ACL2 and
reasoning about it.  Had we taken this route here, the majority of
effort would have been spent on these underlying concepts, precluding
us from examining the procedures of interest quickly and efficiently.
For these reasons, we believe that the encapsulation mechanism was
invaluable in our current work.

\subsection{Constrained Functions and Their Properties}

We constrain four functions using the {\em encapsulate} construct.
The function {\em simplify (x y)} is for simplification of an element  
{\em x} by a set {\em y}.  The function {\em true-symbolp (x)} is a
recognizer for the true symbol (for example, {\em T} or {\em 'true}
or 1) in a particular logic.  The function {\em ceval (x i)} is
for evaluation of a clause {\em x} in interpretation {\em i}. 
The function {\em scount (x)} is for computing the size of the
argument. 

Given witnesses for these four functions, the following constraints
are stated and proved.  Constraints fall into three 
categories depending on which of the three main goals --- termination, 
irredicibility, and logical equivalence --- they enable us to
establish.  To ensure termination of simplification procedures, in
practice we typically use the lexicographic path ordering or the
recursive path ordering \cite{term-rewriting}.   Simplification with
these orderings can increase the number of symbols, so 
{\em acl2-count} does not produce an accurate termination function.
Instead, the constrained function {\em scount} is used to determine
the size of a clause.  The main property of the function is that it
returns a natural number.  

\begin{verbatim}
(defthm scount-natural
  (and (integerp (scount x))
       (<= 0 (scount x))))
\end{verbatim}

\noindent
Termination proofs depend on the constraint that for formulas that are
indeed changed by simplification, the result of the simplification is
somehow smaller than the original expression.
\begin{verbatim}
(defthm scount-simplify
  (or (equal (simplify x y) x)
      (< (scount (simplify x y)) 
         (scount x))))
\end{verbatim}

Proof of the irreducibility property depends on the following
properties of the basic simplification procedure.  An idempotence
property states that once a formula is simplified by a set, attempting
to simplify the result again by the same set will have no effect.
Another property requires that if a set simplifies a formula, then a
superset of that set does so as well.  A third property states that
two sets that do not simplify a formula individually do not do so when
considered collectively.
 
\begin{verbatim}
(defthm simplify-idempotent
  (equal (simplify (simplify x y) y) 
         (simplify x y)))
 
(defthm simplify-subset
  (implies (and (not (equal (simplify a x) a))
                (subsetp-equal x y))
           (not (equal (simplify a y) a))))

(defthm simplify-append
  (implies (and (equal (simplify a x) a)
                (equal (simplify a y) a))
           (equal (simplify a (append x y)) a)))
\end{verbatim}

We formalized the notion of rewritability to improve the readability
of the formalizations of both the direct and limbo incorporation
procedures and to ease management of proofs. If a set simplifies an
element, we say that the element is rewritable by the set.  The new
function {\em rewritable} is defined outside the encapsulation.  Once
the termination and irreducibility constraints are restated in terms
of {\em rewritable}, the function is disabled.

\begin{verbatim}
(defun rewritable (x y)
  (not (equal (simplify x y) x)))
\end{verbatim}

Finally, the proofs of the logical equivalence property of our
incorporation procedures depend on the following properties of the
constrained evaluation function and its relationship with
{\em simplify} and {\em true-symbolp}.  The evaluation function is
Boolean, and the true symbol of the logic is evaluated to true.  We
define a function to evaluate a set of elements as a conjunction.  The
main soundness property of constrained simplification states that if
the conjunction of simplifiers is true, the evaluations of the original
and simplified expressions are equal.

\begin{verbatim}
(defthm ceval-boolean
  (or (equal (ceval x i) t) (equal (ceval x i) nil)))

(defthm true-symbolp-ceval
  (implies (true-symbolp x) (ceval x i)))

(defun ceval-list (x i)
  (if (endp x)
      t
    (and (ceval (car x) i) (ceval-list (cdr x) i))))
 
(defthm simplify-sound
  (implies (ceval-list y i)
           (equal (ceval (simplify x y) i) (ceval x i))))
\end{verbatim}

\subsection{Formalization and Termination of Incorporation Procedures}
Three supporting functions are used to formalize the direct and limbo
incorporation procedures. Rather than present the ACL2 implementation
of the functions, we simply describe them.  The function
{\em extract-rewritables (x s)} computes a subset of elements
of {\em S} that are rewritable by {\em X}.  The function
{\em extract-n-simplify-rewritables (x s)} produces a set of elements
of {\em S} that are rewritable by {\em X} and have been simplified by
it.  The function {\em remove-rewritables (x s)} produces the set of
elements of {\em S} that are not rewritable by {\em X}.  The direct
incorporation procedure is formalized by using the last two functions
as follows.  

\begin{verbatim}
(defun direct-incorporation (q s)
  (cond ((or (not (true-listp q)) (not (true-listp s))) 'INPUT-ERROR)
        ((endp q) s)
        ((true-symbolp (simplify (car q) s)) (direct-incorporation (cdr q) s))
        (t (direct-incorporation 
            (append (cdr q) 
                    (extract-n-simplify-rewritables (simplify (car q) s) s))
            (cons (simplify (car q) s)
                  (remove-rewritables (simplify (car q) s) s))))))
                   
\end{verbatim}

The limbo incorporation procedure relies on computation of the initial
limbo list and subsequent integration of the list into the original
database.  As stated above, the second step of the incorporation
procedure may place new elements on the limbo list.  Before any
element is added to the limbo list, however, it is simplified as much
as possible by the members of the original database and the elements 
already on the limbo list.  We note, therefore, that in the recursive
call of the function {\em preprocess-list}, in addition to the the
members of original database and limbo list, the set of simplifiers
includes elements processed by the function in the previous calls.

\begin{verbatim}
(defun preprocess (x s l)
  (if (true-symbolp (simplify x (append s l)))
      l
    (append l (list (simplify x (append s l))))))

(defun initial-limbo (q s l)
  (if (endp q)
      l
    (initial-limbo (cdr q) s (preprocess (car q) s l))))

(defun preprocess-list (d s r)
  (if (endp d)
      r
      (preprocess-list (cdr d) s (preprocess (car d) 
                                             (append s (cdr d))
                                             r))))

(defun process-limbo (l s)
  (cond ((or (not (true-listp l)) (not (true-listp s))) 'INPUT-ERROR)
        ((endp l) s)
        (t (process-limbo (append (cdr l)
                                  (preprocess-list 
                                   (extract-rewritables (car l) s)
                                   (append (remove-rewritables (car l) s) l)
                                   nil))
                          (cons (car l)
                                (remove-rewritables (car l) s))))))

(defun limbo-incorporation (q s)
  (process-limbo (initial-limbo q s nil) s))
\end{verbatim}

Termination proofs for the functions {\em direct-incorporation} and
{\em process-limbo} rely on the simplification properties stated in
the encapsulation.  The proofs are not entirely trivial; in order to
achieve them, the conjectures must be split into two cases: a case
when the set of elements produced by the {\em extract} functions is
empty, and a case when it is not.  We define an additional counting
function {\em lcount} whose behavior on lists is similar to that of  
{\em acl2-count}, except that the size of list elements is computed
by using the constrained function {\em scount}.

\begin{verbatim}
(defun lcount (x)
  (if (endp x)
      0
     (+ 1 (scount (car x)) (lcount (cdr x)))))
\end{verbatim} 

\noindent 
The measure function, based on {\em lcount}, is

\begin{verbatim}
(cons (+ 1 (lcount q) (lcount s)) 
      (+ 1 (lcount q))).
\end{verbatim} 

\noindent
We note that the formalization on the direct incorporation
procedure is slightly different from the algorithm presented in
Section~\ref{sec:problem}.  In the algorithm elements D that are
rewritable by C are moved from the set S onto Q.  In the
formalization, these elements are simplified by C before being placed
onto Q.  This extra simplification step allows us to show that the
direct incorporation algorithm terminates. Yet this addition to the
original algorithm does not affect the main correctness properties of
the procedure. 

\subsection{Irreducibility Property}
We formulate the irreducibility property as follows.  We
first define a function {\em mutually-irreducible-el-list (x s)} that
checks that the element {\em X} neither rewrites nor is rewritable
by anything in {\em S}.  The main irreducibility check function
relies on the element wise irreducibility check.

\begin{verbatim}
(defun mutually-irreducible-el-list (x s)
  (cond ((endp s) t)
        ((or (rewritable x (list (car s))) 
             (rewritable (car s) (list x))) nil)
        (t (mutually-irreducible-el-list x (cdr s)))))

(defun irreducible-list (s)
  (cond ((endp s) t)
        ((mutually-irreducible-el-list (car s) (cdr s)) 
         (irreducible-list (cdr s)))
        (t nil)))
\end{verbatim}

\noindent
We accomplished the second of the stated goals by proving that if the
original database of clauses is irreducible, both incorporation
procedures produce sets with that property.

\begin{verbatim}
(defthm direct-incorporation-is-irreducible
  (implies (irreducible-list s)
           (irreducible-list (direct-incorporation q s))))

(defthm limbo-incorporation-is-irreducible
  (implies (irreducible-list s)
           (irreducible-list (limbo-incorporation q s))))
\end{verbatim}

\subsection{Soundness}

Soundness proofs rely on the properties of ceval given in the
{\em encapsulate} construct and were relatively easy to establish.  We
showed that both incorporation procedures produce a conjunction of
clauses whose evaluation is equivalent to the evaluation of the
conjunctions of clauses in the two input sets.

\begin{verbatim}
(defthm direct-incorporation-is-sound
  (implies (and (true-listp q)
                (true-listp s))
           (equal (ceval-list (direct-incorporation q s) i)
                  (and (ceval-list q i) (ceval-list s i))))  

(defthm limbo-incorporation-is-sound
  (implies (true-listp s)
           (equal (ceval-list (limbo-incorporation q s) i)
                  (and (ceval-list q i) (ceval-list s i))))     
\end{verbatim}

\section{Related Work and Conclusions}

 Our earlier project {\sc Ivy}
\cite{ivy} dealt with checking the proofs produced by Otter.  The
checker code was written in ACL2 and proved sound.  Although both
efforts concern the same software, the errors they help eliminate do
not overlap.   {\sc Ivy} was designed to catch errors in
Otter-produced proofs.  This work focuses on irreducibility and 
termination, and errors in the simplification procedure
described here would likely not lead to soundness problems in the
resulting proofs, but would prevent Otter from finding some or all
proofs for a particular problem.  

Also related is the large and ongoing ACL2 effort on abstract
reduction systems and term rewriting in \cite{acl2-rewriting-spain}.
The effort concentrates on formalizing basic 
reduction and rewriting procedures in ACL2 and establishing their
properties.  The work includes formalization of first-order logic
and reasoning about termination of rewriting.  Both are aspects that
our effort takes for granted to concentrate on a practical application
that relies on a rewriting procedure.  

The Otter code is based on an algorithm similar to limbo
incorporation.  Correctness of this algorithm is therefore important
to us but is not obvious because of the complexity of the algorithm.
While the algorithm depends on term rewriting and clause subsumption
procedures, we were able, thanks to encapsulation mechanism in ACL2, 
to concentrate on only a few relevant properties of these basic
procedures and to devote all effort to understanding and verifying the
limbo incorporation, the actual procedure of interest. 

\bibliographystyle{plain}
\bibliography{master}

\begin{thebibliography}{1}

\bibitem{term-rewriting}
F.~Baader and T.~Nipkow.
\newblock {\em Term Rewriting and All That}.
\newblock Cambridge University Press, Cambridge, United Kingdom, 1998.

\bibitem{acl2:book}
M.~Kaufmann, P.~Manolios, and J~S. Moore.
\newblock {\em Computer-Aided Reasoning: An Approach}.
\newblock Kluwer Academic Publishers, 2000.

\bibitem{McCune94a}
W.~McCune.
\newblock Otter 3.0 {R}eference {M}anual and {G}uide.
\newblock Tech. Report ANL-94/6, Argonne National Laboratory, Argonne, IL,
  1994.
\newblock See also URL http://www.mcs.anl.gov/AR/otter/.

\bibitem{ivy}
W.~McCune and O.~Shumsky.
\newblock {IVY}: A preprocessor and proof checker for first-order logic.
\newblock In M.~Kaufmann, P.~Manolios, and J~Moore, editors, {\em
  Computer-Aided Reasoning: ACL2 Case Studies}, chapter~16. Kluwer Academic,
  2000.

\bibitem{acl2-rewriting-spain}
J.~L.~Ruiz Reina, J.~A. Alonso, M.~J. Hidalgo, and F.~J. Mart\'{\i}n.
\newblock Formal proofs about rewriting using {A}{C}{L}2.
\newblock {\em Annals of Mathematics and Artificial Intelligence},
  36(3):239--262, 2002.

\end{thebibliography}

\vfill
\begin{flushright}
\scriptsize
\framebox{\parbox{2.4in}{The submitted manuscript has been created
by the University of Chicago as Operator of Argonne
National Laboratory ("Argonne") under Contract No.\
W-31-109-ENG-38 with the U.S. Department of Energy.
The U.S. Government retains for itself, and others
acting on its behalf, a paid-up, nonexclusive, irrevocable
worldwide license in said article to reproduce,
prepare derivative works, distribute copies to the
public, and perform publicly and display publicly, by or on
behalf of the Government.}}
\normalsize
\end{flushright}

\end{document}